# Crocodile perception of distress in Hominid baby cries

Julie Thévenet[1,2], Léo Papet[1,2], Gérard Coureaud[1], Nicolas Boyer[1], Florence Levréro[1], Nicolas Grimault[2‡], Nicolas Mathevon[1,3‡]

[1] ENES Bioacoustics Research Laboratory, CRNL, CNRS, Inserm, University of Saint-Etienne, France;
[2] Equipe Cognition Auditive et Psychoacoustique, CRNL, CNRS, Inserm, University Lyon 1, France;
[3] Institut universitaire de France

‡ These authors are co-last authors

Corresponding author: julie.thevenet.phd@gmail.com

## ABSTRACT

It is generally argued that distress vocalizations, a common modality for alerting conspecifics across a wide range of terrestrial vertebrates, share acoustic features that allow heterospecific communication. Yet studies suggest that the acoustic traits used to decode distress may vary between species, leading to decoding errors. Here we found through playback experiments that Nile crocodiles are attracted to infant hominid cries (bonobo, chimpanzee and human), and that the intensity of crocodile response depends critically on a set of specific acoustic features (mainly deterministic chaos, harmonicity, and spectral prominences). Our results suggest that crocodiles are sensitive to the degree of distress encoded in the vocalizations of phylogenetically very distant vertebrates. A comparison of these results with those obtained with human subjects confronted with the same stimuli further indicates that crocodiles and humans use different acoustic criteria to assess the distress encoded in infant cries. Interestingly, the acoustic features driving crocodile reaction are likely to be more reliable markers of distress than those used by humans. These results highlight that the acoustic features encoding information in vertebrate sound signals are not necessarily identical across species.

# INTRODUCTION

The cross-species perception of emotions conveyed by voice is a topic that tickles both biologists and the general public, as it challenges our ability to understand what non-human animals convey through their vocalizations [1–3]. Darwin had hypothesized that natural selection had led to convergences in emotion coding between animal species, and that the expression of emotions by voice has ancient evolutionary roots. He suggested that the consistency in the way emotions are expressed vocally must be sought in the mechanisms of production by the vocal organs [4]. Indeed, since most airborne vertebrates share the same principles of sound production (vibration of membranes driven by airflow), the acoustic output of emotions is likely to be similar in all these animals, suggesting that they therefore share similar acoustic coding of emotions. A century after Darwin, Morton [5] stated the principle that the acoustic structure of a sound signal should reflect the motivational state of the sender (motivation-structural rules): birds and mammals use rather low-pitched and "rough" sounds in a hostile context, and rather high-pitched and pure tone sounds when they are frightened, or in a friendly context. Emotions can be classified along two dimensions: their arousal level (low to high) and their valence (negative or positive) [6,7]. Recent studies provide empirical data supporting Morton's claim that vocalizations are usually louder and produced at a faster rate, higher pitched, more frequency-modulated and rougher when arousal increases, with positive vocalizations being shorter and less modulated in frequency than negative vocalizations [8–10,1]. Emphasis has been placed on nonlinear phenomena i.e. irregularities in acoustic signals that arise from perturbations of vocal folds vibrations [11]. These nonlinear phenomena (frequency jumps, subharmonics, deterministic chaos and sidebands) [12] are the main cause for the typical roughness of vocalizations related to high arousal. The arousal level of a distress call is accordingly correlated with the amount of acoustic non-linearity [13–18]. Nonlinear acoustic phenomena are therefore good candidates to encode distress in the sound signals across animal species.

If there are commonly shared coding rules, it can be hypothesized that airborne vertebrates - including humans- are able to decode emotions conveyed by heterospecific acoustic signals. For instance, people who live with pets often claim to be able to decipher emotions such as distress, joy or surprise, by listening to their companion meow or bark [19–21]. Livestock and pet professionals often agree [22–24], and they even seem to be able to decode more finely than non-professionals the vocalizations of domestic animals (see the recent paper by Massenet et al. on the perception of puppy whines by humans [14]). However, individual impressions, while suggesting the existence of inter-species communication, do not provide objective information about the processes of decoding the emotions carried by acoustic signals. A scientific approach, based on acoustic analyses of emotion-carrying signals and on playback experiments to identify the acoustic parameters relevant to the receivers, is needed to shed light on the processes involved. Indeed, and although a number of studies have supported this hypothesis by showing, for example, that human subjects are able to estimate the degree of arousal in the calls of different vertebrate species [2,21,25–28], other studies suggest however that the acoustic criteria used by receivers are not always appropriate. For example, Kelly et al. [29] tested how human adult listeners rate the level of distress expressed in the distress calls (cries) of the infants of human and non-human apes (bonobo and chimpanzee). They showed that human listeners pay attention to the mean pitch of vocalizations, leading them to overestimate the level of distress encoded in the bonobo cries. The very high-pitched bonobo infant calls were indeed systematically rated as expressing overall high distress levels despite being recorded in contexts eliciting various stress intensity. Similarly, Teichrob et al. [30] and Lingle and Riede [31] find that female mule deer and white-tailed deer respond to distress calls from different mammals only when the frequency range of these calls is artificially brought into the frequency range of baby deer vocalizations. Root-Gutteridge et al. [32] also showed that the response of domestic dogs to the cries of puppies and human babies is highly dependent on their frequency range. Thus, the ability to identify vocally encoded emotions across species may be driven by species-specific traits [33]. Furthermore, we have very limited knowledge of how non-human animals can decode emotions in the vocalizations of other animal species since most studies focus on the perception of animal vocalizations by humans. Moreover, investigations in non-human animals usually involved species that are phylogenetically close (e.g., cross-species perception of alarm and distress calls in birds [34]; deer perception of distress calls from other mammals [31]).

Crocodilians and hominids (great apes including humans) are distant vertebrate groups on the phylogenetic tree. However, both crocodile and hominid infants solicit parental care through distress vocalizations [35–37] (hereafter called "cries"). Caregivers -a parent in most species- are attracted to these infant cries and respond by providing various types of parental care (protection from predators in crocodiles; protection and feeding in hominids). Moreover, crocodiles are top predators and sometimes commit cannibalism [38,39]. As they are opportunistic feeders, crocodilians use all their senses to detect and locate their preys [39]. A such, they can be attracted to the cries of potential prey of many species[39][39][39]. Crocodile hearing abilities are excellent between 100 Hz and 4 kHz [35], and it is empirically known that the whines of puppies and the cries of human babies attract crocodiles [40]. Crocodiles are therefore an excellent model for investigating an animal's ability to identify distress in the vocalizations of other phylogenetically distant species. Furthermore, it is noteworthy that birds and mammals are about the only non-human animals that have been tested for the perception of human vocal features. Crocodilians are closely related to birds and have many ecological and cognitive traits that converge with mammals. Although we have a fairly good knowledge of their anatomy, their behavior -and in particular their behavioral responses to acoustic stimuli- still remains poorly known [41].

The present study extends this comparative approach, testing whether crocodilians react to distress vocalizations of phylogenetically distant species, the cries of infant hominids. We identify the specific acoustic features modulating the responses of Nile crocodiles to these distress calls, and we compare these "crocodilian" acoustic features to what we had found previously with humans [29]. We predict that these acoustic features will be identical to those that modulate the assessment of distress level encoded in vocalizations in humans, which would suggest that the decoding of vocal emotions in acoustic signals could be shared by phylogenetically distant species. We first conducted an analysis of the acoustic structure of hominid baby cries (bonobos, chimpanzees and humans) recorded in different situations eliciting various levels of distress. We then observed the behavioral response of adult Nile crocodiles to the cries and identified the acoustic traits of the cries that explain the variation in reaction intensity of the crocodiles. Finally, we compared these results with those obtained in adult humans in a previous study using the same stimuli[29]. We found that, unlike adult humans whose response to baby cries is primarily driven by cry frequency pitch, crocodiles are particularly attentive to a set of features including nonlinear acoustic phenomena, which are known to be particularly informative of distress level encoded in vocalizations of several mammals [13].

## METHODS

*Acoustic stimuli*

All stimuli have been recorded prior to the present study and are part of the sound database of the ENES Bioacoustics Research Laboratory (Figure 1A).

Bonobo infant cries were recorded in European zoological parks, with a Sennheiser MKH70 ultra-directional microphone connected to a Zoom H4n recorder (sampling frequency = 44.1 kHz). The recorded infants were aged from 1 to 4 years (sex unknown) and all dependent on their mother (breast-fed and frequently carried by the mother). We isolated 6 cries from 6 different bonobo babies (one cry per baby; cry duration = 3.0 ± 0.4 seconds).

Chimpanzee infant cries were recorded in the wild from a population habituated to humans (Kibale National Park, Uganda), with Sennheiser MKH70 ultra-directional microphone connected to a Marantz PMD670 digital recorder (sampling frequency = 44.1 kHz). The exact ages as well as the sex of the recorded infants were unknown, but all were under 4 years and were carried by their mother.

We isolated 6 cries from 6 different chimpanzee babies (one cry per baby; cry duration = 3.2 ± 0.4 seconds).

The context of the bonobo and chimpanzee infant cries recordings was always characterized by an infant soliciting its mother. For practical and ethical reasons, it was not possible to quantify the emotional state of the bonobo and chimpanzee infants by physiological measures (e.g., heart rate).

An operational definition is to consider arousal as a marker of the intensity of an emotion, ranging from low to high [1]. In this framework, calls caused by minor stress correspond to low arousal, while calls caused by major stress represent high arousal. We thus arbitrarily classified the cries of chimpanzee and bonobo babies into 2 categories: low arousal (begging calls while the mother was nearby, without a third individual involved in the interaction; n = 3 bonobo cries, n = 2 chimpanzee cries), and high arousal (conflict with another individual, no immediate reaction from the mother; n = 3 bonobo cries, n = 4 chimpanzee cries). However, given the limited number of bonobo and chimpanzee recordings of respectively low and high arousal, we chose to group them into a single category to perform the acoustic analysis. Importantly for the purposes of the study, we obtained recordings of bonobo and chimpanzee infants soliciting their mothers in various situations. This gave our cry sample a wide range of arousal levels which allows us to explore the acoustic basis of the distress response in crocodiles.

Human babies' cries were recorded in two contexts: bathing at home by parents (low arousal context; N = 6 babies with balanced sex ratio) and vaccination at the pediatrician's office (high arousal context; N = 6 babies with balanced sex ratio, different from those recorded in the bathing context). The babies were recorded with a Sennheiser MD42 microphone placed at 30 cm from their mouth, and connected to a Zoom H4n recorder (sampling frequency = 44.1 kHz). We isolated one cry per recorded baby (total of 12 sequences; duration of each cry = 3.2 ± 0.2 seconds).

From a general point of view, the bank of cry recordings with which we worked thus contained a diversity of signals, both in terms of hominid species and in terms of arousal level. This diversity allowed us to achieve the major objective of the study, which was to identify the specific acoustic features modulating the response of crocodiles to distress calls.

### Analysis of sound stimuli

We analysed the acoustic structure of cry sequences using a custom script in *Praat* software [42]. Extending the analysis method we developed previously [29], we measured the following eighteen acoustic variables: number of cry syllables in the sequence (***nbCries***), average duration of each syllable in the sequence (***meanDur***), percentage of the sequence duration with detectable pitch (*voiced*), average pitch over the sequence (*meanF0*), minimum pitch (*minF0*), maximum pitch (*maxF0*), range of pitch (maxF0 – minF0 = *rangeF0*), pitch coefficient of variation (*F0CV*), harmonicity (*harmonicity*; ratio of harmonics to noise in the signal expressed in dB), jitter index (*jitter*; small fluctuations of periodicity), index of shimmer (*shimmer*; small variations in amplitude), first three spectral prominences characterizing the spectral envelope of the cries (*SP1*, *SP2*, *SP3*), percentage of the sequence duration with subharmonics (*subharmonics*; nonlinear phenomena appearing on the spectrogram as integer fractional values of an F0), percentage of the sequence duration with biphonation (*biphonation*; nonlinear phenomena characterized by two simultaneous and independent fundamental frequencies), percentage of the sequence duration with deterministic chaos (*chaos*; nonlinear phenomena characterized by non-random noise), mean intensity of the cry sequence (*meanINTcroc* or *meanINThuman*). The latter variable was calculated by considering the respective auditory sensitivity of crocodiles and humans. The maximum amplitude of all signals was previously normalized. Based on the crocodilian audiogram measured by Higgs et al. (2002), we converted flat dB into "dB crocodile" and obtain the average intensity corresponding to the hearing sensitivity of crocodiles (see Supplementary Material 1 for details). The average human intensity was calculated using dB(A).

To illustrate the differences between cry categories while reducing the acoustic dimensions considered, we performed a principal component analysis (PCA) on all 18 acoustic variables (taking *meanINTcroc* as the measure for the mean intensity of the cry sequence). Clustering of cry categories according to the first three principal components (acoustic dimensions) was tested using anova (package *stats*, R-Studio v.4.1.2), taking each acoustic dimension as the dependent variable and the cry category (bonobo, chimpanzee, human baby at bath, human baby during vaccination) as the fixed factor. We then performed post hoc multiple comparisons of means (Tukey contrasts, R package *multcomp*).

### *Playback experiments on crocodiles*

The experiments were conducted at CrocoParc zoo (Agadir, Morocco). This park hosts more than 300 Nile crocodiles (*Crocodylus niloticus*) in an outdoor garden including several ponds. As the animals are free to roam, it was impossible to test the crocodiles individually. We therefore played back the sound stimuli to four groups of between 7 and 25 adults (females and males, unknown sex ratio), occupying 4 different ponds (see Figure 1A for a plan of the ponds and the position of the speakers). By emitting acoustic stimuli to groups of crocodiles rather than to single individuals, the possibility that individual responses to the stimuli are influenced by the behavior of other animals in the basin cannot be excluded. To our knowledge, no previous study has experimentally tested that crocodiles mimic the behavior of other individuals - although they are probably capable of doing so (e.g., some species, such as the spectacled caiman, black caiman and alligator, have been shown to fish in groups [44]). Above all, in the basins where we did the experiments, the crocodiles are rarely on the move, and when one or a few of them start to swim, they are never systematically followed by others. The fact that we observed, as in this study, rapid responses towards the speaker (see results below) argues in favor of reactions driven first by the sound stimuli. In order to avoid habituation, each group of crocodiles was tested during a single experimental session, except for one group that was tested in two sessions, one day apart. To accustom the crocodiles to the presence of the speakers, they were positioned two days before the start of the experiments. During these two days, we regularly observed the crocodiles (several observation sessions per day) and noticed that they showed no particular interest in these silent speakers. Each experimental session started at 19:00, one hour after the park closed to the public.

During each experimental session, each group of crocodiles heard a succession of up to 6 stimuli (min-max = 4-6), broadcasted with a remote-controlled loudspeaker (FoxPro Fusion with Visaton SL 87 ND internal speakers, see Supplementary Figure 1 for the technical specifications). Each stimulus consisted of a 30-second repetition of one of the 24 previously isolated cries. The stimuli were different between groups. Among the stimuli sent to each group, there was at least one cry from each category (bonobo, chimpanzee, human baby bathing, human baby being vaccinated). To avoid any order effect, we also took the precaution of presenting the stimuli in different orders and combinations from one experiment to another (see Supplementary Table 1 for a detailed list of stimuli played to the different groups). There was a minimum interval of 10 min between each playback, and we only played the next stimulus when the crocodiles had lost interest in the speaker.

All experiments were filmed (Lumix DMC-FZ300 camera). The behavioural response of the crocodiles was assessed by measuring the proportion of individuals who responded to the stimulus (number of individuals who turned their head toward the speaker or moved in its direction divided by the number of individuals present in the basin during the experiment).

### Analysis of the crocodile reaction to playback

We tested the effect of the cry category on crocodile behavioral response with a generalized mixed model (package *lme4*, logistic function, R-Studio v.4.1.2, fixed factor: cry category, random factor: number of the pond where the tested group of crocodiles was located; the individual identity of the recorded crying infants was not taken into account because 22 out of 24 stimuli were played only once). We used multiple comparison tests to compare the behavioral responses of crocodiles across stimulus categories.

To identify the acoustic traits that could explain crocodiles behavioral responses in terms of response rate (i.e. proportion of crocodiles reacting to the sound stimulus), we explored the relative importance of each of the 18 acoustic variables characterizing the stimuli using partial least squares logistic regression PLS (package *plsRglm* [45]). PLS is useful when a response has to be predicted from a large set of variables and when there is multicollinearity. While classical principal component analysis does not identify the salient acoustic features explaining the behavioral response, PLS-regression allows to group the acoustic features that best predict these responses. PLS constructs components from linear combinations of the predictors optimized to be related to the variable to be explained. Here, the variable to be explained was the crocodile response rate to the acoustic stimuli while the predictors were the 18 acoustic parameters. Cross validation was used to select the optimal number of components in the model. Predictor significance and BCa confidence intervals were derived using balanced bootstrap (R=1000 resampling). Results were expressed as standardized regression coefficients $\beta$ and credible intervals derived from the bias corrected accelerated bootstrap distribution. Coefficients with bootstrap distributions above or below zero were considered statistically significant. Finally, we tested the potential correlation between the value of the PLS first component and the amount of distress expressed by the cries.

### *Comparison with humans*

The perception of distress encoded in infant bonobo, chimpanzee, and human cries by adult human listeners has been the subject of the previous study by Kelly et al. [29]. However, in order to compare with the results obtained here with crocodiles, we reanalyzed the data from that previous study normalizing human ratings between 0 and 1 and using the acoustic variables presented here. We conducted the same analyses as described above for crocodiles (principal component analysis and partial least squares logistic regression, taking *meanINThuman* instead of *meanINTcroc*).

## RESULTS

### *Cry stimuli differ by pitch, presence of chaos and distribution of energy in the spectrum*

Principal Component Analysis performed on the 18 acoustic variables yields three major components (called "Acoustic Dimensions") that significantly discriminate the cry categories (anova, AD1: $F(3,20) = 70.6$, $p < 0.001$; AD2: $F(3,20) = 22.5$, $p < 0.001$; AD3: $F(3,20) = 8.4$, $p < 0.001$). The first acoustic dimension (AD1) primarily represents cry pitch (with *meanF0*, *maxF0*, and *rangeF0* showing the highest loadings; Table 1). As illustrated by Figure 1B, bonobo cries have a higher pitch than those of chimpanzees and human babies. Human babies recorded during bathing were the lowest pitched cries (see Table 2 for post-hoc multiple comparisons between cry categories). The second acoustic dimension AD2 essentially represents deterministic chaos (nonlinear phenomenon) and the highest spectral prominences (SP2 and SP3). It is the cries of human babies recorded in a vaccination context and the cries of bonobos that present the highest values of AD2 (Figure 1B). The third dimension (AD3) is associated with the periodic quality of the cries (captured by the variables *voiced*, *harm* and *jitter*; Figure 1B). As shown in Figure 1B, chimpanzee and particularly bonobo calls are distributed along a gradient spread along all three acoustic dimensions, meaning that our recordings can be considered representative of acoustic variability in the cries of these great apes.

### *Crocodile response is driven by a set of acoustic features that do not include pitch*

The results of the playback experiments show no significant difference between the response rates of crocodiles to different distress cry categories (GLM, Wald $X^2$ = 5.0, p = 0.173). As suggested in Figure 2A, only some high-arousal human infant cries and some bonobo cries appear to stand out by inducing a stronger response than low-arousal cries or chimpanzee cries. However, the response of crocodiles is dependent on the acoustic characteristics of the stimuli. Indeed, the PLS regression reveals the acoustic predictors of crocodile response to sound stimuli (Figure 2B). Low harmonicity, high jitter, the presence of chaos, and higher energy in the higher frequencies of the spectrum accompany higher responsiveness of the tested animals (harmonicity: β = -0.16 [-0.43, -0.09]; jitter: β = 0.14 [0.07, 0.39]; deterministic chaos: β = 0.11 [0.03, 0.21]; SP2: β = 0.13 [0.07, 0.36]; SP3: β = 0.09 [0.03, 0.19]). Conversely, the pitch (F0) does not predict crocodile reaction to sound stimuli. Only one component was kept in the PLS model, and as illustrated in Figure 2C, this first component (Crocodile PLS1, Supplementary Table 2) is significantly higher (lmer, Wald $X^2$ = 8.6, p < 0.01) when crocodiles were tested with signals potentially expressing a lot of distress (high arousal cries), than when they hear signals potentially expressing less distress (low arousal cries).

### *Human listeners rely mostly on pitch features to assess baby cries*

Similar to the results reported in [29], we found that human listeners judge the cry of hominid babies differentially (GLM, Wald $X^2$ = 92.0, p < 0.001, Figure 2C). Specifically, they rate bonobo cries as expressing the highest level of distress (multiple comparisons: β > 1.0, Z > 3.5, p < 0.01), while human babies' low arousal cries are rated as expressing the least distress (multiple comparisons: β < -1.2, Z < -5.4, p < 0.001).

PLS regression reveals that the most significant predictors are related to pitch and its variation (rangeF0: β = 0.08 [0.07, 0.09]; maxF0: β = 0.07 [0.06, 0.08]; F0CV: β = 0.08 [0.06, 0.10]). Human listeners thus assign a high distress value to high-pitched cries. Other predictors, such as harmonicity (β = -0.07 [-0.08, -0.06]), also modulate human listeners' rating of cries (Figure 2D).

Similarly to crocodiles, the first component of the PLS model (Human PLS1, Supplementary Table 2) is higher when individuals are tested with signals expressing a lot of distress than when they hear signals expressing less distress (Figure 2F).

## DISCUSSION

Our results show that Nile crocodiles are attracted to infant hominid cries and suggest that their motivation to respond depends on acoustic features known to encode the intensity of distress expressed by the emitter. In particular, crocodiles are more attracted to cries with nonlinear acoustic phenomena (chaos, low harmonicity) and more intense energy in the high frequencies of the spectrum (spectral prominences), which are two acoustic traits known to code for a high arousal. Unlike humans who primarily use pitch to judge the level of distress encoded in infant cries, crocodiles are only moderately sensitive to this acoustic feature.

The present analysis of the acoustic structure of bonobo, chimpanzee, and human infant cries confirms and complements the work of Kelly et al. [29]. The acoustic parameter differing the most between stimulus categories is pitch. Bonobo infant cries are by far the highest pitched, followed by human infant high arousal cries. Human baby high arousal cries are distinguished from other stimuli by a greater presence of nonlinear phenomena [28,46] (deterministic chaos, low harmonicity). As illustrated by the representation of the stimuli in two-dimensional acoustic space, some of the bonobo cries also exhibit marked nonlinear phenomena [47] (high values of the second acoustic dimension). Because of their diversity, our stimuli thus represent a range of distress coding. Human infants were recorded in two contexts (at the bath and during a vaccination session) inducing different levels of arousal. Bonobo and chimpanzee infants were recorded in a variety of contexts, also inducing a diversity of arousal levels.

This diversity in the acoustic features of the stimuli, which reflects a diversity in the arousal of the emitters, drives a variation in the behavioral response of the crocodiles. Playback experiments suggest that crocodiles are not particularly sensitive to the category of the playback call (bonobo, chimpanzee or human), but pay particular attention to acoustic features that mark arousal, such as nonlinear acoustic phenomena (chaos and low harmonicity) and spectral prominences in the upper part of the spectrum. Our psychoacoustic approach allows to explore the correlations between the objective acoustic properties of sound stimuli and the subjective perception of sounds by crocodiles. The results of the acoustic analysis show that pitch is an essential dimension for differentiating sounds in the acoustic space determined by principal component analysis. On the other hand, the results of the playback experiments highlight that pitch is not a reliable predictor of the behavioural response of crocodiles. In short, what we can call the "perceptual sound dimension", i.e., the acoustic dimension that best explains the crocodiles' response, does not correspond exactly to the acoustic dimensions that best discriminate sound stimuli.

What might be termed the "perceptual dimension" of crocodiles does not correspond exactly to the distribution of calls in the acoustic space determined by principal component analysis. In particular, while it is an essential dimension for discriminating between cry categories, pitch is not a reliable predictor of crocodile response.

This result is interesting for two reasons. First, it marks a difference with the way humans assess the level of distress in infant cries. Our analysis of humans' assignment of distress level to stimuli indeed confirms that conducted by Kelly et al. [29]: humans assign a distress level primarily from the pitch of the cry. The higher the pitch of a cry, the more humans judge the cry as expressing high distress [27,21,25,2]. This leads human listeners to consistently judge bonobo baby cries as expressing strong distress, and thus to be likely to be wrong. Bonobo babies have been recorded in a variety of contexts and express a diversity of arousal levels, which is reflected in the wide distribution of their cries on the second acoustic dimension (chaos and spectral prominences). The second benefit of crocodiles not paying attention to pitch is that their reaction to the baby cry is less dependent on the animal species emitting the cry than it is for humans. Pitch is a potentially misleading distress marker trait: while it may be informative within a given animal species, its basal value varies too much between animal species to be a marker of distress widely shared across species [33,32,29–31]. Crocodiles thus seem particularly adapted at estimating the degree of distress encoded in an infant's cry regardless of the hominid species considered. Unlike humans, whose perception and analysis of infant cries is biased by this emphasis on pitch, crocodiles probably have no experience with cries in different hominid species - except in the wild for crocodiles living in close proximity to human populations or other Hominids. They therefore respond to stimuli based on acoustic criteria alone, without recognizing the origin of the stimulus. Moreover, although crocodiles have excellent hearing in air, they [35,43] have a poor perception of high frequencies and bonobo cries, which have a frequency range up to 10 kHz, are therefore likely to be poorly perceived by crocodiles [35,43]. This poor perception of high frequencies may explain why human babies' high arousal cries tend to induce a stronger response from crocodiles.

Why humans rely heavily on the pitch (F0) to assess the level of distress coded by a cry is a question beyond the scope of this paper. To put it in a nutshell, this is potentially explained by the fact that a human infant expresses a high level of distress by increasing the prevalence of not only nonlinear acoustic phenomena but also pitch [28]. The average pitch differs between human babies, but since a human normally knows the baby(s) they are caring for, using pitch to detect distress in a given baby's cries becomes reliable.

Why do crocodiles respond to hominid cries? It is known that adult crocodiles are attracted to the distress calls of their young. Crocodilian females -and males in some species- come to provide parental assistance in case of attack by a predator [48]. However, adult crocodilians do not always show care for the young, and cases of cannibalism are regularly observed [38,39]. The distress calls of young crocodilians share common features with the cries of hominid infants: they are harmonic series spanning a wide bandwidth (between 500 and 6000 Hz) [49,50], modulated in frequency, and may exhibit nonlinear acoustic phenomena. In our experiments, it was not possible to reliably identify each tested adult crocodile as male or female. However, we definitely observed that individuals of both sexes responded to our stimuli. In addition, not all animals approached the speaker in the same way. Some did so by swimming on the surface, while others practiced an underwater approach, a behavior which looks as a predator strategy. We also observed some individuals trying to bite the speaker, while others stopped a few tens of centimeters in front of it with their snout pointed in its direction. Although it is likely that our hominid cry stimuli triggered a predatory response from these opportunistic hunters, it cannot be entirely ruled out that some individuals (particularly females) responded in a parental care context.

Furthermore, by showing that adult Nile crocodiles are attracted to the cries of hominid babies, our study suggests that these large predators may have been a danger to the human lineage throughout its evolution. The Nile crocodile was indeed an abundant species in the African cradle where the human lineage developed [51]. Since the cries of babies of all species forming the human lineage probably shared acoustic characteristics with the cries of present-day human babies, they probably always represented attractive stimuli for crocodiles.

One possible limitation of our study is the lack of a control stimulus. Indeed, we tested the crocodiles only with cries, and not with any other type of signal. However, we are confident that the responses we observed are indeed driven by cries and not by any other type of sound. In fact, our previous works with crocodilians show that these animals respond selectively to certain sound signals. For example, a black caiman mother does not respond to her offspring's contact calls while she rushes to a speaker emitting distress calls [50]. Young Nile crocodiles are attracted to calls from other young, but do not respond to frog calls [52,53]. In the present study, we could certainly have played non-significant signals (such as white noise for example). We chose not to do so in order to limit the number of playback experiments. Crocodiles are particularly intelligent animals and their ability to associate the presence of a speaker with listening to unusual sounds is certainly very high. Since the number of testable animals was very limited, we needed to keep the number of experiments to a minimum.

Since our stimuli were distress calls recorded from young individuals, it is not possible to generalize our results to other emotional vocalizations, such as adult distress calls. Only new playback experiments could answer this question. For instance, it would be interesting to test the reaction of crocodiles to other types of vocalizations, including vocalizations with a positive valence such as laughter. Based on the results of the present study, we hypothesize that since these vocalizations show little or no non-linear phenomena, they should elicit less response from the crocodiles.

In sum, our study suggests that crocodiles identify and respond proportionately to different levels of distress in hominid infant cries. This behavior is elicited by acoustic features otherwise known to be markers of distress in vocalizations [13–15,54,16,55,18]. Our experiments obviously do not mean that crocodiles cannot be attracted by other signals than distress calls -they are opportunistic hunters-, but they suggest that the readiness of these animals to react increases with the presence of acoustic features marking a level of distress (non-linear acoustic phenomena). Moreover, and because the crocodiles tested have never had the opportunity to associate a hominid baby's cry with the presence of a baby in their immediate surroundings in the zoo, their response to hominid baby cries is likely innate. Since crocodiles primarily pay attention to nonlinear acoustic phenomena where humans are primarily sensitive to pitch, our results suggest that the decoding of arousal level expressed by vocalizations may be based on different acoustic features depending on the species.

## DATA ACCESSIBILITY

All datasets and R scripts required for reproducing the results presented in the paper will be uploaded in a Zenodo repository with a dedicated Digital Object Identifier (DOI).

## ETHICS

All experiments were performed in accordance with relevant guidelines and regulations in Morocco and under the supervision of zoo's staff, and were approved by the Institutional Animal Ethical Com-

## AUTHORS' CONTRIBUTIONS

JT: conceptualization, data acquisition, analysis, writing – original draft and writing – editing; LP: data acquisition, writing – editing; GC: data acquisition, writing – editing; NB: data acquisition; FL: recording of stimuli; NG: conceptualization, data acquisition, analysis, writing – editing, NM: conceptualization, data acquisition, writing – editing.

## ACKNOWLEDGEMENTS

The authors are grateful to "Crocoparc" zoo in Agadir, Morocco (Luc Fougeirol, Ariane Marinetti, Leila Sdigui, as well as Philippe and Christine Alleon and the staff). This research has been funded by the Institut universitaire de France (NM), the Labex CeLyA (Lyon Center of Acoustics ANR-10-LABX-60, PhD funding to Julie Thévenet and Léo Papet), the CNRS and the University of Saint-Etienne.

## COMPETING INTERESTS

The authors declare no competing interests.

## REFERENCES

1. Briefer EF. 2020 Coding for 'Dynamic' Information: Vocal Expression of Emotional Arousal and Valence in Non-human Animals. In *Coding Strategies in Vertebrate Acoustic Communication* (eds T Aubin, N Mathevon), pp. 137–162. Cham: Springer International Publishing. (doi:10.1007/978-3-030-39200-0_6)

2. Filippi P *et al*. 2017 Humans recognize emotional arousal in vocalizations across all classes of terrestrial vertebrates: evidence for acoustic universals. *Proceedings of the Royal Society B: Biological Sciences* **284**, 20170990. (doi:10.1098/rspb.2017.0990)

3. Scheumann M, Hasting AS, Kotz SA, Zimmermann E. 2014 The Voice of Emotion across Species: How Do Human Listeners Recognize Animals' Affective States? *PLOS ONE* **9**, e91192. (doi:10.1371/journal.pone.0091192)

4. Darwin C. 1871 *The Descent of Man, and Selection in Relation to Sex*. Princeton University Press. (doi:10.1515/9781400820061)

5. Morton ES. 1977 On the Occurrence and Significance of Motivation-Structural Rules in Some Bird and Mammal Sounds. *The American Naturalist* **111**, 855–869. (doi:10.1086/283219)

6. Laukka P, Juslin P, Bresin R. 2005 A dimensional approach to vocal expression of emotion. *Cognition and Emotion* **19**, 633–653. (doi:10.1080/02699930441000445)

7. Mendl M, Burman OHP, Paul ES. 2010 An integrative and functional framework for the study of animal emotion and mood. *Proceedings of the Royal Society B: Biological Sciences* **277**, 2895–2904. (doi:10.1098/rspb.2010.0303)

8. August PV, Anderson JGT. 1987 Mammal Sounds and Motivation-Structural Rules: A Test of the Hypothesis. *Journal of Mammalogy* **68**, 1–9. (doi:10.2307/1381039)

9. Ehret G. 2006 Common rules of communication sound perception. *Behavior and Neurodynamics for Auditory Communication* , 85–114.

10. Briefer EF. 2012 Vocal expression of emotions in mammals: mechanisms of production and evidence. *Journal of Zoology* **288**, 1–20. (doi:10.1111/j.1469-7998.2012.00920.x)

11. Fitch WT, Neubauer J, Herzel H. 2002 Calls out of chaos: the adaptive significance of nonlinear phenomena in mammalian vocal production. *Animal Behaviour* **63**, 407–418. (doi:10.1006/anbe.2001.1912)

12. Wilden I, Herzel H, Peters G, Tembrock G. 1998 Subharmonics, Biphonation, and Deterministic Chaos in Mammal Vocalization. *Bioacoustics* **9**, 171–196. (doi:10.1080/09524622.1998.9753394)

13. Lingle S, Wyman M, Kotrba R, Teichroeb L, Romanow C. 2012 What makes a cry a cry? A review of infant distress vocalizations. *Current Zoology* **58**, 698–726. (doi:10.1093/czoolo/58.5.698)

14. Massenet M, Anikin A, Pisanski K, Reynaud K, Mathevon N, Reby D. 2022 Nonlinear vocal phenomena affect human perceptions of distress, size and dominance in puppy whines. *Proceedings of the Royal Society B: Biological Sciences* **289**, 20220429. (doi:10.1098/rspb.2022.0429)

15. Anikin A, Pisanski K, Massenet M, Reby D. 2021 Harsh is large: nonlinear vocal phenomena lower voice pitch and exaggerate body size. *Proceedings of the Royal Society B: Biological Sciences* **288**, 20210872. (doi:10.1098/rspb.2021.0872)

16. Marx A, Lenkei R, Pérez Fraga P, Bakos V, Kubinyi E, Faragó T. 2021 Occurrences of non-linear phenomena and vocal harshness in dog whines as indicators of stress and ageing. *Sci Rep* **11**, 4468. (doi:10.1038/s41598-021-83614-1)

17. Blumstein DT, Richardson DT, Cooley L, Winternitz J, Daniel JC. 2008 The structure, meaning and function of yellow-bellied marmot pup screams. *Animal Behaviour* **76**, 1055–1064. (doi:10.1016/j.anbehav.2008.06.002)

18. Stoeger AS, Charlton BD, Kratochvil H, Fitch WT. 2011 Vocal cues indicate level of arousal in infant African elephant roars. *The Journal of the Acoustical Society of America* **130**, 1700–1710. (doi:10.1121/1.3605538)

19. Prato-Previde E, Cannas S, Palestrini C, Ingraffia S, Battini M, Ludovico LA, Ntalampiras S, Presti G, Mattiello S. 2020 What's in a Meow? A Study on Human Classification and Interpretation of Domestic Cat Vocalizations. *Animals (Basel)* **10**, E2390. (doi:10.3390/ani10122390)

20. Faragó T, Takács N, Miklosi A, Pongracz P. 2017 Dog growls express various contextual and affective content for human listeners. *Royal Society Open Science* **4**, 170134. (doi:10.1098/rsos.170134)

21. Faragó T, Andics A, Devecseri V, Kis A, Gácsi M, Miklósi Á. 2014 Humans rely on the same rules to assess emotional valence and intensity in conspecific and dog vocalizations. *Biology Letters* **10**, 20130926. (doi:10.1098/rsbl.2013.0926)

22. Laurijs KA, Briefer EF, Reimert I, Webb LE. 2021 Vocalisations in farm animals: A step towards positive welfare assessment. *Applied Animal Behaviour Science* **236**, 105264. (doi:10.1016/j.applanim.2021.105264)

23. Manteuffel G, Puppe B, Schön PC. 2004 Vocalization of farm animals as a measure of welfare. *Applied Animal Behaviour Science* **88**, 163–182. (doi:10.1016/j.applanim.2004.02.012)

24. Tallet C, Špinka M, Maruščáková I, Šimeček P. 2010 Human perception of vocalizations of domestic piglets and modulation by experience with domestic pigs (Sus scrofa). *Journal of Comparative Psychology* **124**, 81–91. (doi:10.1037/a0017354)

25. Maruščáková IL, Linhart P, Ratcliffe VF, Tallet C, Reby D, Špinka M. 2015 Humans (Homo sapiens) judge the emotional content of piglet (Sus scrofa domestica) calls based on simple acoustic parameters, not personality, empathy, nor attitude toward animals. *Journal of Comparative Psychology* **129**, 121–131. (doi:10.1037/a0038870)

26. Pongrácz P, Molnár C, Miklósi Á. 2006 Acoustic parameters of dog barks carry emotional information for humans. *Applied Animal Behaviour Science* **100**, 228–240. (doi:10.1016/j.applanim.2005.12.004)

27. McComb K, Taylor AM, Wilson C, Charlton BD. 2009 The cry embedded within the purr. *Current Biology* **19**, R507–R508. (doi:10.1016/j.cub.2009.05.033)

28. Koutseff A, Reby D, Martin O, Levrero F, Patural H, Mathevon N. 2018 The acoustic space of pain: cries as indicators of distress recovering dynamics in pre-verbal infants. *Bioacoustics* **27**, 313–325. (doi:10.1080/09524622.2017.1344931)

29. Kelly T, Reby D, Levréro F, Keenan S, Gustafsson E, Koutseff A, Mathevon N. 2017 Adult human perception of distress in the cries of bonobo, chimpanzee, and human infants. *Biol J Linn Soc* **120**, 919–930. (doi:10.1093/biolinnean/blw016)

30. Teichroeb LJ, Riede T, Kotrba R, Lingle S. 2013 Fundamental frequency is key to response of female deer to juvenile distress calls. *Behavioural Processes* **92**, 15–23. (doi:10.1016/j.beproc.2012.09.011)

31. Lingle S, Riede T. 2014 Deer Mothers Are Sensitive to Infant Distress Vocalizations of Diverse Mammalian Species. *The American Naturalist* **184**, 510–522. (doi:10.1086/677677)

32. Root-Gutteridge H, Ratcliffe VF, Neumann J, Timarchi L, Yeung C, Korzeniowska AT, Mathevon N, Reby D. 2021 Effect of pitch range on dogs' response to conspecific vs. heterospecific distress cries. *Sci Rep* **11**, 19723. (doi:10.1038/s41598-021-98967-w)

33. Bremond J-C. 1976 Specific recognition in the song of Bonelli's warbler (*Phylloscopus bonelli*). *Behaviour* **58**, 99–116. (doi:10.1163/156853976X00253)

34. Aubin T. 1991 Why do distress calls evoke interspecific responses? An experimental study applied to some species of birds. *Behavioural Processes* **23**, 103–111. (doi:10.1016/0376-6357(91)90061-4)

35. Vergne A, Pritz M, Mathevon N. 2009 Acoustic communication in crocodilians: From behaviour to brain. *Biological reviews of the Cambridge Philosophical Society* **84**, 391–411. (doi:10.1111/j.1469-185X.2009.00079.x)

36. Bard KA. 2000 Crying in infant primates: Insights into the development of crying in chimpanzees. In *Crying as a sign, a sympton, & a signal: Clinical emotional and developmental aspects of infant and toddler crying*, pp. 157–175. New York, NY, US: Cambridge University Press.

37. Bermejo M, Omedes A. 2000 Preliminary Vocal Repertoire and Vocal Communication of Wild Bonobos (Pan paniscus) at Lilungu (Democratic Republic of Congo). *Folia Primatologica* **70**, 328–357. (doi:10.1159/000021717)

38. Hutton J. 1989 Movements, Home Range, Dispersal and the Separation of Size Classes in Nile Crocodiles. *American Zoologist* **29**, 1033–1049.

39. Grigg GC, Kirshner D. 2015 *Biology and evolution of crocodylians*. Ithaca: Comstock Publishing Associates a division of Cornell University Press.

40. Silva A de, Somaweera R. 2015 Were human babies used as bait in crocodile hunts in colonial Sri Lanka? *Journal of Threatened Taxa* **7**, 6805–6809. (doi:10.11609/JoTT.o4161.6805-9)

41. Reber SA. 2020 Crocodilians Are Promising Intermediate Model Organisms for Comparative Perception Research. *Comparative Cognition & Behavior Reviews* **15**, 111–129. (doi:10.3819/CCBR.2020.150004)

42. Boersma P, Weenink D. 1992 PRAAT: Doing phonetics by computer (Version 6.2.06).

43. Higgs D, Brittan-Powell E, Soares D, Souza M, Carr C, Dooling R, Popper A. 2002 Amphibious auditory responses of the American alligator (Alligator mississipiensis). *Journal of comparative physiology. A, Neuroethology, sensory, neural, and behavioral physiology* **188**, 217–23. (doi:10.1007/s00359-002-0296-8)

44. King F, Thorbjarnarson J, Yamashita C. 1998 Cooperative feeding, a misinterpreted and under-reported behavior of crocodilians. *Florida Museum of Natural History* , 9.

45. Bertrand F, Maumy M, Meyer N. 2009 plsRglm, modèles linéaires généralisés PLS sous R. In *Chimiométrie 2009*, p. pp 52-54. Paris, France.

46. Raine J, Pisanski K, Simner J, Reby D. 2019 Vocal communication of simulated pain. *Bioacoustics* **28**, 404–426. (doi:10.1080/09524622.2018.1463295)

47. Waal FBMD. 1988 The Communicative Repertoire of Captive Bonobos (Pan Paniscus), Compared To That of Chimpanzees. *Behaviour* **106**, 183–251. (doi:10.1163/156853988X00269)

48. Somaweera R, Brien M, Shine R. 2013 The Role of Predation in Shaping Crocodilian Natural History. *Herpetological Monographs* **27**, 23. (doi:10.1655/HERPMONOGRAPHS-D-11-00001)

49. Vergne A, Avril A, Martin S, Mathevon N. 2007 Parent-offspring communication in the Nile crocodile Crocodylus niloticus: Do newborns' calls show an individual signature? *Die Naturwissenschaften* **94**, 49–54. (doi:10.1007/s00114-006-0156-4)

50. Vergne AL, Aubin T, Taylor P, Mathevon N. 2011 Acoustic signals of baby black caimans. *Zoology* **114**, 313–320. (doi:10.1016/j.zool.2011.07.003)

51. Fergusson RA, Faso B, Guinea E. 2010 Nile Crocodile Crocodylus niloticus. *Crocodiles : status survey and conservation action plan* , 84–89.

52. Vergne AL, Aubin T, Martin S, Mathevon N. 2012 Acoustic communication in crocodilians: information encoding and species specificity of juvenile calls. *Anim Cogn* **15**, 1095–1109. (doi:10.1007/s10071-012-0533-7)

53. Thévenet J, Kehy M, Boyer N, Pradeau A, Papet L, Gaudrain E, Grimault N, Mathevon N. 2023 Sound categorization by crocodilians. *iScience* **26**, 106441. (doi:10.1016/j.isci.2023.106441)

54. Anikin A, Pisanski K, Reby D. 2020 Do nonlinear vocal phenomena signal negative valence or high emotion intensity? *Royal Society Open Science* **7**, 201306. (doi:10.1098/rsos.201306)

55. Rendall D, Notman H, Owren MJ. 2009 Asymmetries in the individual distinctiveness and maternal recognition of infant contact calls and distress screams in baboons. *The Journal of the Acoustical Society of America* **125**, 1792–1805. (doi:10.1121/1.3068453)

56. American National Standards Institute, United States of America Standards Institute. 1967 *Preferred frequencies and band numbers for acoustical measurements*. New York: United States of America Standards Institute.

# Tables

**Table 1: Principal Component Analysis of the sound stimuli.**

| Acoustic parameters | AD1 Variance = 39.6% Eigenvalue = 7.1 | AD2 Variance = 20.4% Eigenvalue = 3.7 | AD3 Variance = 12.2% Eigenvalue = 2.2 |
|---:|---|---|---|
| nbCries | 0.35 | -0.65 | -0.33 |
| meanDuration | -0.44 | 0.69 | -0.07 |
| voiced | 0.08 | -0.05 | **0.72** |
| meanF0 | **0.91** | -0.17 | 0.05 |
| maxF0 | **0.94** | -0.16 | 0.18 |
| minF0 | 0.31 | -0.51 | -0.31 |
| rangeF0 | **0.91** | -0.04 | 0.27 |
| F0CV | 0.42 | 0.44 | 0.48 |
| meanINTcroc | -0.68 | -0.41 | -0.32 |
| harmonicity | -0.62 | -0.27 | **0.54** |
| jitter | 0.77 | 0.08 | **-0.54** |
| shimmer | 0.79 | -0.21 | -0.43 |
| SP1 | 0.70 | 0.41 | -0.03 |
| SP2 | 0.47 | **0.75** | -0.03 |
| SP3 | 0.48 | **0.74** | -0.12 |
| biphonation | 0.84 | -0.14 | 0.32 |
| subharmonics | -0.37 | 0.22 | -0.03 |
| chaos | -0.35 | **0.79** | -0.40 |

Table 2: Multiple comparisons tests between sound stimuli.

| Cry categories comparison | Acoustic Dimension 1 | | | Acoustic Dimension 2 | | | Acoustic Dimension 3 | | |
|---|---|---|---|---|---|---|---|---|---|
| | Estimate | t | p | Estimate | t | p | Estimate | t | p |
| Human bath / Human vaccine | -1.66 | -3.36 | **0.015** | -3.57 | -6.16 | **< 0.001** | 2.18 | 3.49 | **0.011** |
| Human bath / bonobo | -6.68 | -13.48 | **< 0.001** | -1.48 | -2.55 | 0.081 | 0.03 | 0.05 | 1.000 |
| Human bath / chimpanzee | -4.34 | -8.76 | **< 0.001** | 0.87 | 1.50 | 0.454 | 2.29 | 3.67 | **< 0.01** |
| Human vaccine / bonobo | -5.02 | -10.12 | **< 0.001** | 2.09 | 3.61 | **< 0.01** | -2.15 | -3.44 | **0.013** |
| Human vaccine / chimpanzee | -2.68 | -5.40 | **< 0.001** | 4.44 | 7.66 | **< 0.001** | 0.11 | 0.18 | 1.000 |
| Bonobo / chimpanzee | 2.34 | 4.72 | **< 0.001** | 2.35 | 4.06 | **< 0.01** | 2.26 | 3.62 | **< 0.01** |

**Figures legends**

**Figure 1. Playback experiments with Nile crocodiles. A.** Aerial view of the ponds of *Crocoparc zoo*, Agadir, Morocco. We played the sound stimuli to 4 groups of 7 to 25 adults (females and males, sex ratio unknown), occupying 4 different ponds. For each experimental session, two loudspeakers were placed on the banks of the pond, allowing to playback the sound stimuli from two different locations. Arrows indicate the position and direction of the loudspeakers. Camera icons indicate the position of the cameras used to monitor the experimental trials. **B.** Hominid baby cries used as playback stimuli. Top: Spectrograms of cry samples. Bottom: Acoustic space of cries. Each dot represents a cry stimulus (duration around 3 seconds). The acoustic structure of cries was described using 18 acoustic variables and further reduced into three independent acoustic dimensions using a Principal Component Analysis. The first acoustic dimension is mainly related to cry pitch. The second acoustic dimension is mainly related to deterministic chaos and high spectral prominences. The third acoustic dimension is mainly related to cry harmonicity (see Table 1 for Principal Components coefficients). Legend of dots in the acoustic spaces: disks = low-arousal human baby cries; triangles = high arousal human baby cries; empty squares = low arousal baby bonobo cries; solid squares = high arousal baby bonobo cries; simple crosses = low arousal baby chimp cries; double crosses = high arousal baby chimp cries.

**Figure 2: Response of crocodile and human adults to hominid baby cries. A.** Crocodile response rates to sound stimuli. Each group of crocodiles was tested with all stimuli in several successive experimental trials. Each dot represents an experimental trial. The outcome measured was the proportion of responding individuals in each tested crocodile group. **B.** PLS Acoustic predictors of crocodile response to baby cries. Coefficients with bootstrap distributions above or below zero are statistically significant predictors. The crocodile response to cries is essentially predicted by low harmonicity, jitter, the presence of chaos and the highest spectral prominences. Predictors related to stimulus pitch do not explain the crocodile response (Partial Least Square Regression: standardized regression coefficients and 95% balanced bias-corrected and accelerated bootstrap confidence intervals). **C.** Relationship between the PLS first component and the degree of distress expressed by cries in the experiment involving crocodiles. **D.** Distress level ratings of sound stimuli in the experiment involving human adults. Each dot represents one psychoacoustic measure. **E.** PLS Acoustic predictors of human rating of distress level in baby cries. Coefficients with bootstrap distributions above or below zero are statistically significant predictors. Predictors related to stimulus pitch and its variation explain the best human ratings (Partial Least Square Regression: standardized regression coefficients and 95% balanced bias-corrected and accelerated bootstrap confidence intervals). **F.** Relationship between the PLS first component and the degree of distress expressed by cries in humans.

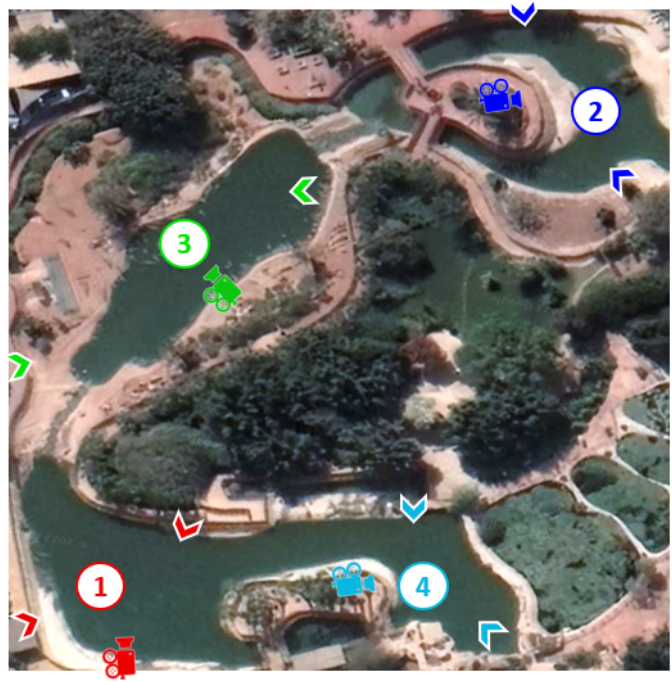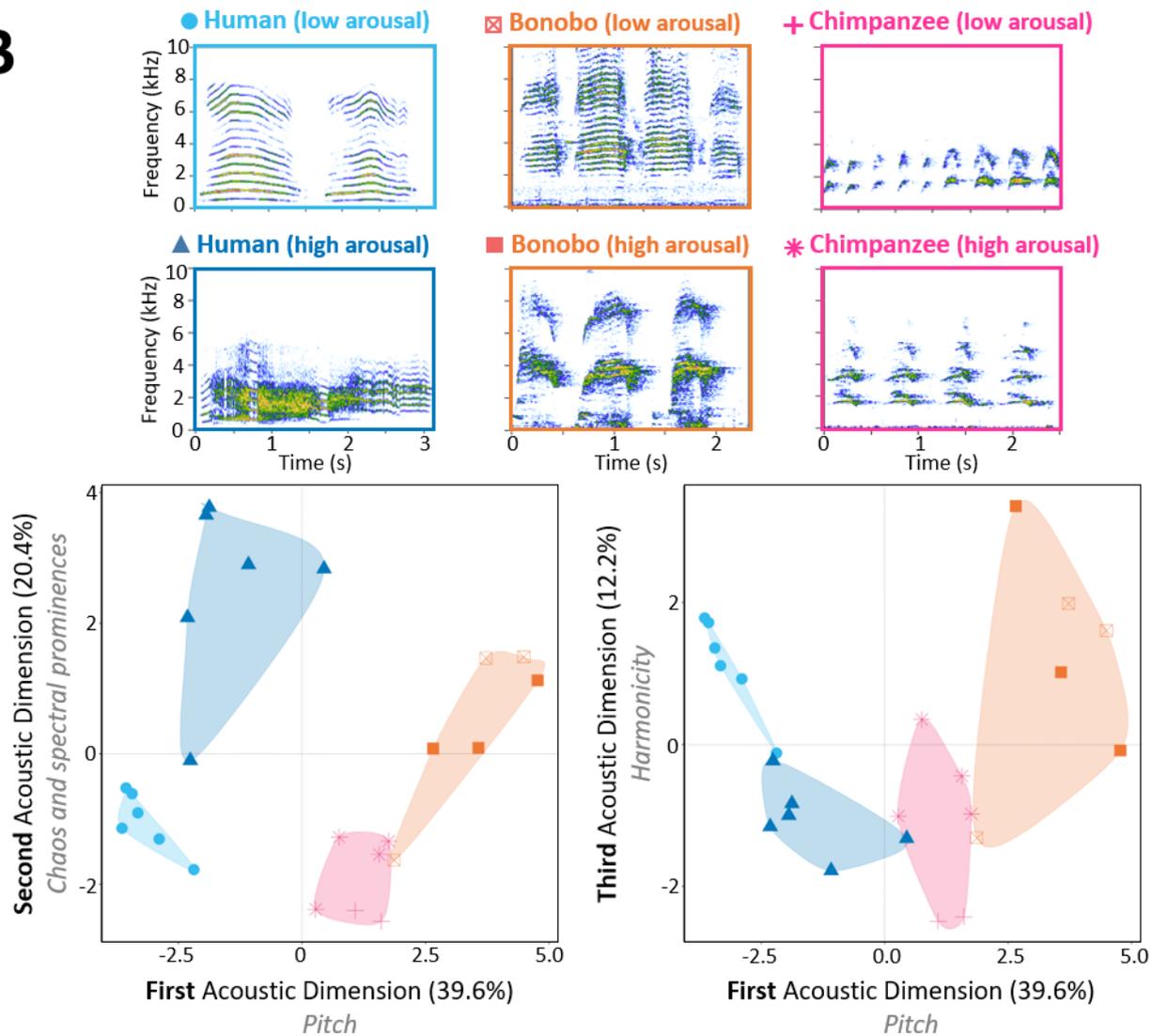

# Crocodile

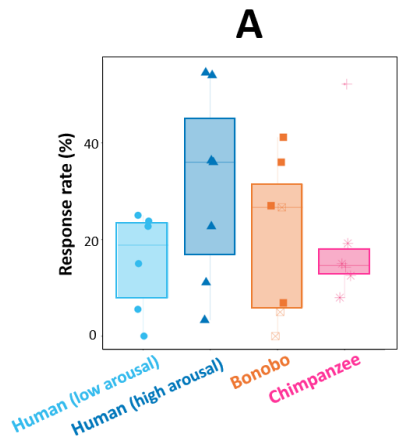 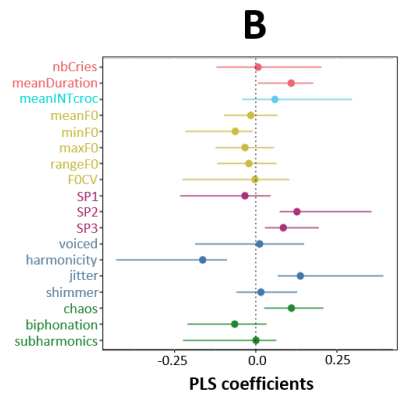 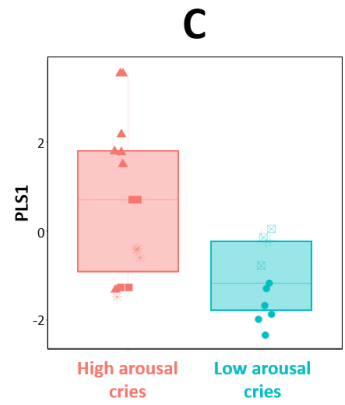

**A**     **B**     **C**

# Human

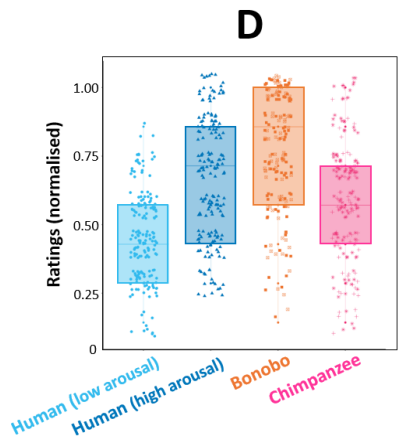 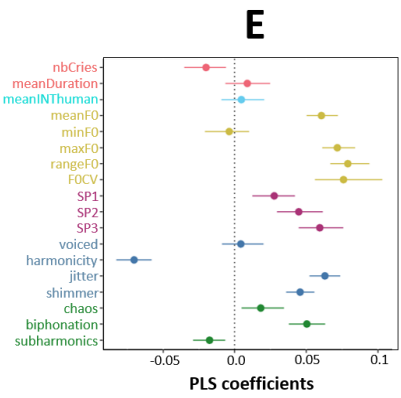 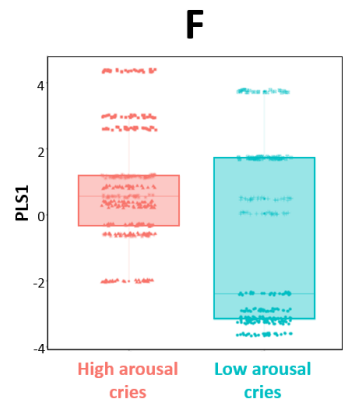

**D**     **E**     **F**

# Supplementary Materials

**Supplementary Material 1: Conversion of flat dB to dB "crocodile".** Based on the ANSI methodology used to estimate the sound level in dB(A) (i.e. related to the human auditory curve; ANSI S1.6-1967 (R1976)) [56], we developed the same methodology based on the alligator audiogram as measured by Higgs et al. (2002) to estimate the sound intensity in dB "crocodile". The first step was to interpolate a weighting function in dB for each frequency from the alligator audiogram curve (ref 0 dB at 1000Hz) and to transform this weighting function on a linear scale. The second step was to compute the spectrum of the sounds in dB using an FFT (Fast Fourier Transform) and then to multiply this FFT transformed on linear scale by the weighting function. Finally, we calculated the mean RMS level in dB of the resulting FFT.

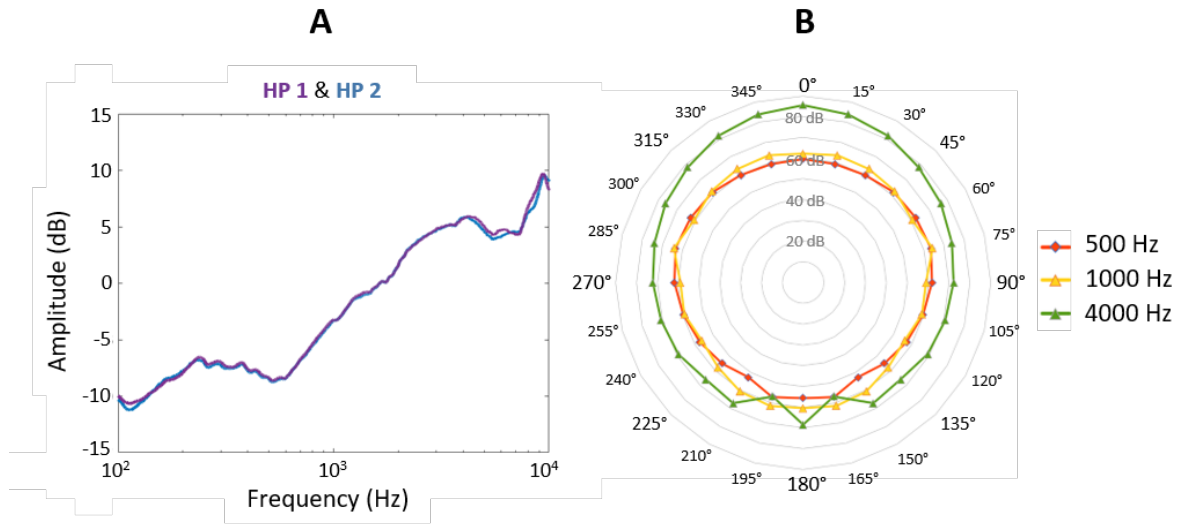

**Supplementary Figure 1. Technical specifications of the loudspeakers.** Most of the energy of the acoustic signals used in this study was between 500 Hz and 10 kHz. Since the original transfer functions of the FoxPro loudspeakers (Shogyo #GF0923BM-1X) in this frequency range were very hilly with rapid variations up to 16 dB, we decided to insert a new speaker into the FoxPro with smoother and flatter transfer functions (Visaton #SL 87 ND). In addition, the use of these new loudspeakers makes the FoxPro much more omnidirectional. All measurements have been done in a semi-anechoic room (dimensions 3.4 x 4.6 m; reverberation time = 0.2 for 125 Hz and ≤ 0.1 for frequencies greater than or equal to 500 Hz). **A.** Transfer functions of the two FoxPro (1 and 2) with the new loudspeakers measured with a condenser microphone (Behringer ECM8000) while emitting a broadband noise. **B.** Directionality functions (polar diagram, dB scale) of the FoxPro with the new intern speakers measured for sinus at three frequencies.

**Supplementary Table 1: Details of the playback experiments.**

| Pond | Trial | Cry category | Signal ID | Nb of individuals in the pond | Response rate |
|---|---|---|---|---|---|
| 1 | 1 | bonobo | 03_bonobo (high arousal) | 25 | 36 % |
| 1 | 2 | chimpanzee | 04_chimp (high arousal) | 20 | 15 % |
| 1 | 3 | Human bath | 01_human_bath | 18 | 6 % |
| 1 | 4 | bonobo | 03_bonobo (high arousal) | 22 | 27 % |
| 1 | 5 | Human vaccine | 02_human_vaccine | 11 | 55 % |
| 1 | 6 | Human vaccine | 02_human_vaccine | 28 | 54 % |
| 2 | 1 | Human bath | 06_human_bath | 7 | 0 % |
| 2 | 2 | Human vaccine | 11_human_vaccine | 11 | 36 % |
| 2 | 3 | bonobo | 16_bonobo (low arousal) | 12 | 0 % |
| 2 | 4 | chimpanzee | 21_chimp (high arousal) | 8 | 13 % |
| 3 | 1 | chimpanzee | 22_chimp (low arousal) | 23 | 52 % |
| 3 | 2 | Human vaccine | 12_human_vaccine | 30 | 3 % |
| 3 | 3 | bonobo | 17_bonobo (low arousal) | 30 | 27 % |
| 3 | 4 | Human bath | 07_human_bath | 21 | 24 % |
| 3 | 5 | bonobo | 18_bonobo (high arousal) | 29 | 7 % |
| 3 | 6 | chimpanzee | 23_chimp (low arousal) | 21 | 14 % |
| 4 | 1 | Human bath | 08_human_bath | 20 | 25 % |
| 4 | 2 | bonobo | 19_bonobo (high arousal) | 17 | 41 % |
| 4 | 3 | Human vaccine | 13_human_vaccine | 22 | 23 % |
| 4 | 4 | chimpanzee | 24_chimp (high arousal) | 25 | 8 % |
| 4 | 5 | Human bath | 09_human_bath | 20 | 15 % |
| 4 | 6 | Human vaccine | 14_human_vaccine | 18 | 11 % |
| 1 | 1 | Human vaccine | 15_human_vaccine | 25 | 36 % |
| 1 | 2 | chimpanzee | 25_chimp (high arousal) | 26 | 19 % |
| 1 | 3 | bonobo | 20_bonobo (low arousal) | 20 | 5 % |
| 1 | 4 | Human bath | 10_human_bath | 22 | 23 % |

**Supplementary Table 2: Weights of predictor variables on the first PLS component.**

| Acoustic parameters | Crocodile PLS1 | Human PLS1 |
|---:|:---:|:---:|
| nbCries | 0.02 | -0.10 |
| meanDuration | **0.33** | 0.04 |
| voiced | 0.04 | 0.02 |
| meanF0 | -0.05 | **0.30** |
| maxF0 | -0.10 | **0.35** |
| minF0 | -0.19 | -0.02 |
| rangeF0 | -0.06 | **0.39** |
| F0CV | -0.01 | **0.37** |
| meanINTcroc | 0.18 | 0.02 |
| harmonicity | **-0.50** | **-0.34** |
| jitter | **0.42** | 0.31 |
| shimmer | 0.05 | 0.22 |
| SP1 | -0.10 | 0.13 |
| SP2 | **0.39** | 0.22 |
| SP3 | 0.26 | 0.29 |
| biphonation | -0.20 | 0.25 |
| subharmonics | 0.002 | -0.09 |
| chaos | **0.33** | 0.09 |